\documentclass[sigconf,american]{acmart}
\usepackage[utf8]{inputenc}
\usepackage{lmodern}
\usepackage{mathrsfs}
\usepackage[T1]{fontenc}
\usepackage{babel}
\usepackage{relsize}            %
\usepackage{xspace}             %
\usepackage[inline,shortlabels]{enumitem}
\usepackage{url}
\usepackage{nameref}
\usepackage{xtab}
\usepackage{booktabs}  %
\usepackage{longtable}
\usepackage{array}
\usepackage{tabularx}
\usepackage{multirow}
\usepackage{siunitx}   %
\usepackage{u8chars}   %
\usepackage{listings}  %
\usepackage{tikz}
\usepackage[textsize=footnotesize,textwidth=1.5cm]{todonotes}
\usepackage{moreverb}
\usepackage[normalem]{ulem}	%

\newcommand{\cf}		{{\upshape	      cf.}\xspace}

\newcommand{\eg}		{{e.g.}\xspace}

\newcommand{\ie}		{{i.e.}\xspace}

\newcommand{\wrt}		{{w.r.t.}\xspace}

\newcommand{\inote}[2][]{\todo[inline,color=green!20,caption={2do}, #1]{%
    \begin{minipage}{\textwidth-4pt}#2\end{minipage}}}
\newcommand{\itodo}[2][]{\todo[inline,caption={2do}, #1]{%
    \begin{minipage}{\textwidth-4pt}#2\end{minipage}}}

\newenvironment{redenv}{\color{red}}{}
\newcommand{\red}[1]{{#1}}

\newcommand\PathCrawler{{\sffamily PathCrawler}\xspace}
\newcommand\Klee{{\sffamily Klee}\xspace}
\newcommand\LLVM{{\sffamily LLVM}\xspace}
\newcommand\LTest{{\sffamily LTest}\xspace}
\newcommand\LAnnot{{\sffamily LAnnotate}\xspace}

\newcommand\FramaC{{\sffamily \mbox{Frama-C}}\xspace}
\newcommand\clang{{\sffamily clang}\xspace}
\newcommand\Kleelab{{\sffamily Klee4labels}\xspace}
\newcommand\pclabel{label\xspace}
\newcommand\pclabels{labels\xspace}
\newcommand\DC{\textbf{DC}\xspace}
\newcommand\loc{\ensuremath{\mathit{loc}}\xspace}

\newcommand\criterionName{\(\mathscr{C}\)\xspace}
\newcommand\nan{\textbf{\color[rgb]{0.57, 0.36, 0.51}{—}}\xspace}
\newcommand\timeout{{\color[rgb]{0.57, 0.36, 0.51}{t.o.}}\xspace}

\lstdefinestyle{inlined}{%
  breakatwhitespace,%
  basicstyle=\ttfamily\small,%
}
\makeatletter
\newcommand\lstatenable[1][]{%
  \protect\lstMakeShortInline[style=inlined,mathescape,#1]@%
}
\makeatother

\setcopyright{acmcopyright}
\acmDOI{10.1145/3555776.3577713}

\acmISBN{978-1-4503-9517-5/23/03}

\acmConference[SAC'23]{ACM SAC Conference}{March 27 –April 2, 2023}{Tallinn, Estonia}
\acmYear{2023}
\copyrightyear{2023}

\acmArticle{12927.130}
\acmPrice{15.00}

\makeatletter%
\begin{document}

\title{An Efficient Black-Box Support of Advanced Coverage Criteria for \Klee}

\def\OCamlPro{%
  \institution{OCamlPro}
  \city{Paris}
  \country{France}
}

\author{Nicolas Berthier}
\author{Steven De Oliveira}
\affiliation{%
  \OCamlPro
}
\email{nicolas.berthier@ocamlpro.com}
\email{steven.de-oliveira@ocamlpro.com}

\author{Nikolai Kosmatov,
	Delphine Longuet,
	Romain Soulat}
\affiliation{%
  \institution{Thales Research \& Technology}
  \city{Palaiseau}
  \country{France}
}
\email{nikolai.kosmatov@thalesgroup.com}
\email{delphine.longuet@thalesgroup.com}
\email{romain.soulat@thalesgroup.com}

\renewcommand{\shortauthors}{N. Berthier et al.}

\begin{abstract}
Dynamic symbolic execution (DSE) is a powerful test generation approach
based on an exploration of the path space of the program under test.
Well-adapted for path coverage, this approach is however
less efficient for conditions, decisions, advanced coverage criteria
(such as multiple conditions, weak mutations, boundary testing)
or user-provided test objectives.
While theoretical solutions to adapt DSE to a large set of criteria
have been proposed, they have never
been integrated into publicly available testing tools.
This paper presents a first integration of an optimized
test generation strategy for advanced coverage criteria
into a popular open-source testing tool based on DSE, namely, \Klee.
The integration is performed in a fully black-box manner, and can therefore
inspire an easy integration into other similar tools.
The resulting version of the tool, \red{named \Kleelab}, is publicly available.
We present the design of the proposed technique and evaluate it on several
benchmarks. Our results confirm the benefits of the proposed tool
for advanced coverage criteria.
\end{abstract}



\lstatenable                    %

\maketitle

\section{Introduction}
\label{sec:intro}

Automatic test generation techniques have made significant
progress during the past two decades.
One of the most remarkable successes in that area is \emph{dynamic symbolic execution} (DSE)~\cite{CadarGKPSTV11SEPrelAssess,CadarS13SEThreeDecades},
an efficient test generation technique combining symbolic and concrete executions of the program under test.
Several efficient DSE tools have been developed~\cite{GodefroidKS05DART,SenMA05CUTE,WMMR05,CadarGPDE06EXE,CDE08,TH08,SageGodefroidLM08,PR10,TH08},
and several case studies and industrial applications have been reported
\cite{CadarGPDE06EXE,CDE08,SageGodefroidLM08,BKMMW18}.

Dynamic symbolic execution relies on an
exhaustive exploration of the path space of the program under test.
Depending on the tool, several strategies can be available:
depth-first search, breadth-first search, as well as
more elaborated heuristics.
While DSE is well-adapted for path coverage,
it is less suitable for other coverage criteria,
such as conditions, decisions, multiple conditions, weak mutations,
boundary testing, or user-provided test objectives.
Indeed, various coverage criteria require to cover quite different
test objectives in the program code, and the support of different criteria
in the existing tools remains limited.
While the all-path criterion is powerful, it is sometimes too
expensive when only decisions or conditions need to be covered.
And the other way round, the tests generated for the all-path criterion may
miss interesting behaviors with respect to such criteria as
multiple conditions, weak mutations, or limit values. In
practice, validation engineers aim at satisfying a given coverage
criterion with a test suite of manageable size.

\citet{BKC14} proposed a framework for specifying coverage criteria
in a unified way
with elementary test objectives,  called \emph{(coverage) labels},
and proposed theoretical solutions for an efficient test generation
for labels.
However, these techniques have never been integrated into any
publicly available DSE tool.
The only known implementation of these techniques is
a greybox integration~\cite{BKC14} inside \PathCrawler~\cite{WMMR05},
a proprietary test generation tool,
whose industrial evaluation was reported to be
quite efficient~\cite{BKMMW18}.

The use of labels to represent coverage criteria allows
for a very generic approach to criteria-guided test generation.
Labels provide a lightweight solution to encode various criteria, which is
entirely independent of the underlying test generation tool.
Furthermore, many basic and advanced coverage criteria
can be represented using labels, and even custom labels can
be added by hand when specific test objectives are needed.
The absence of a publicly available DSE tool with an efficient support
for a large range of coverage criteria remains
a serious barrier for a larger support of coverage criteria.
Addressing this issue is the main motivation of this work.

Our goal is to demonstrate how a dedicated support for labels
can be integrated into \Klee~\cite{CDE08}, a popular open-source test
generation tool based on DSE.
Contrary to the earlier greybox integration into \PathCrawler~\cite{BKC14},
whose path exploration strategy
had to be strongly modified and adapted for labels,
we perform a fully black-box integration,
that can inspire a lightweight integration into other similar tools.
\red{The resulting version of the tool, named \Kleelab, is publicly available.}
Finally, we evaluate the proposed version on several benchmarks.
Overall, those benchmarks demonstrate the ability of the proposed tool:
\begin{enumerate*}[(i)]
\item to drive symbolic execution towards the test objectives required by a given coverage criterion,
\item to generate fewer and more meaningful test cases,
\item with a reasonable overhead compared to the original version of the tool.
\end{enumerate*}

\paragraph{Contributions.}
The contributions of this work include:
\begin{itemize}
\item
a lightweight black-box integration of an efficient support for various coverage
criteria expressed by labels in a popular publicly available tool \Klee, making it
possible to support a large panel of coverage criteria;
\item
a detailed description of the integration as a generic approach that is
expected to be reproducible with other similar tools;
\item
an evaluation of the extended version of \Klee on several benchmarks, and a detailed
analysis of their results in comparison with other approaches, confirming
the benefits of the proposed technique.
\end{itemize}

\paragraph{Outline.}
Section~\ref{sec:background} presents necessary background on \red{dynamic symbolic execution}, the \Klee tool, and on expressing
test coverage criteria with coverage labels.
Some motivating examples are given in Section~\ref{sec:motiv-ex}.
The design of our optimized test generation technique for \Klee is presented in Section~\ref{sec:instru}.
Section~\ref{sec:experiments} provides a detailed evaluation of the proposed technique
in comparison to other approaches for dealing with labels.
Related work is discussed in Section~\ref{sec:related}.
Finally, Section~\ref{sec:conclusion} gives a conclusion and some directions for future work.

\vspace{-2mm}
\section{Background}
\label{sec:background}

\subsection{Dynamic Symbolic Execution}

\red{Symbolic execution~\cite{King76,CadarS13SEThreeDecades} is a powerful approach to automatic test
  generation, based on systematic path exploration. It consists in
  computing, for each path of the program under test, its \emph{path
  condition}, that is, the set of conditions on the program
  parameters that must hold to ensure that the program executes along this
  path (under the assumption that the program is deterministic). If
  this path condition is satisfiable, then a solution gives concrete
  values to the parameters, and thus a test case that executes this
  path. If it is not satisfiable, then the path is called infeasible,
  meaning that it does not correspond to any possible execution of the
  program. In practice, the path may contain statements about which it
  is difficult to reason symbolically, for example, calls to external
  functions, which can lead to an under-approximation of the path
  constraint. Dynamic symbolic execution~\cite{CadarGKPSTV11SEPrelAssess} interleaves concrete and
  symbolic execution and uses the gathered information to better
  approximate path constraints.}  %

\vspace{-2mm}
\subsection{\Klee}
\label{sec:klee}

\Klee~\cite{CDE08} is a popular test-case generation tool for C programs
based on {dynamic symbolic execution}, developed and maintained
at Imperial College London.
\Klee is open-source, has a large community of contributors, and a large user base, both in industry and in academia.
\Klee operates on \LLVM bitcode, which is an intermediate representation of the executable program: this enables it to mix concrete and symbolic executions.
\Klee internally makes use of various {(configurable)} strategies to explore the path space of the program under test, and is able to produce a test case for each path found to be feasible.
Such a test case consists of concrete input values on which the program executes along this path.
By design, \Klee's only coverage criterion is
\emph{all-path}: as a result, the user often must configure a timeout or specify preconditions to the program in order to ensure \Klee's termination within reasonable time bounds.
More precisely, \Klee aims at covering all paths of the LLVM bitcode, which means in particular that compound conditions are decomposed according to lazy evaluation semantic of Boolean operators in C.

Concretely, when launched on a given program, \Klee tries to produce a test case for each executable path that reaches a return statement of the main function,
an assertion, or an instruction that may raise a runtime error (RTE), \eg  division by 0, over\-shift, invalid pointer access.
A path that reaches a return statement is called a \emph{complete path}, whereas other paths are called \emph{partially completed paths}.
In \Klee's output directory, a test case generated for a partially completed path comes with an additional file that ends with \texttt{.\(\mathit{xxx}\).err}, where \(\mathit{xxx}\) gives information about the premature end of the path, \eg \texttt{assert} (for assertion failure), or \texttt{ptr} (for pointer error).
\Klee can replay generated test cases in a separate step.

Since \Klee aims at covering all paths of the LLVM bitcode, it sometimes produces more test cases than needed to cover only decisions or conditions for example.
Its path-oriented approach
is not directly adapted to more advanced coverage criteria like multiple conditions, boundary tests or weak mutations (\cf Section~\ref{sec:coverage-labels}).
The purpose of this work is to
adapt and optimize \Klee
for a large range of coverage criteria expressed using coverage labels.

\vspace{-2mm}
\subsection{Coverage Labels}
\label{sec:coverage-labels}

\citet{BKC14} introduce a generic approach to represent coverage criteria as source code annotations.
They propose to represent the test objectives required to be covered to satisfy a given
coverage criterion as a set of annotations named \emph{coverage labels}
(that we often abbreviate as \emph{labels}).
A coverage label ℓ is defined as a pair $(\loc,p)$, where $p$ is a predicate attached to some program location \loc.
Such a label is covered by a test
if the execution of this test reaches
the location \loc and satisfies the predicate $p$ at this location.
In this way, when the underlying test generation tool
produces test cases covering all labels corresponding to a given coverage criterion,
it builds a test suite satisfying the corresponding criterion.
Previous work proposed \LTest~\citep{BNMD21}, a toolset dedicated to labels.
It is developed as a set of plugins of \FramaC~\citep{KKP15},
a verification platform for C code.
The \LTest toolset notably comprises the \LAnnot tool that,
given a C program and a coverage criterion,
automatically annotates that program by adding the corresponding labels.
More precisely, given a coverage criterion \criterionName
and the C source code of a program \(P\), it automatically outputs a
C program \(P'\) with additional coverage labels with respect to \criterionName, \ie
such that a test suite satisfies \criterionName iff it covers all the added labels.

Basic coverage criteria can be simulated by coverage labels: for example, instruction coverage (\textbf{IC}), decision coverage (\textbf{DC}) or condition coverage (\textbf{CC}, where all atomic conditions of each decision, and their negations, must be covered).
Each label encodes an elementary test objective required to be covered.

Coverage labels can additionally encode more advanced criteria on conditions, boundary (that is, limit) values or mutations.
Multiple condition coverage (\textbf{MCC}) requires to cover all combinations of truth values for all atomic conditions of each decision.
For a chosen set of mutation operators,
weak mutation coverage (\textbf{WM}) requires to cover  (or, as it is often called, to \emph{kill})
each mutant program.
A mutant is killed by a test execution if the mutation point is reached and,
after the mutated instruction,
some variable has a different value in the mutant compared to the original program.
This can typically be encoded by a label with a predicate ensuring non-equality
between the original and mutated expressions.
If such a label is covered by a test, this test kills the corresponding
mutant (in the aforementioned sense).
Many common weak mutation operators, such as ABS (absolute value insertion), ROR (relational operator replacement), AOR (arithmetic operator replacement), and COR (logical operator replacement), can be simulated with labels.

Boundary testing requires to cover limit values of variables or conditions.
For example, in the condition limit coverage criterion,
for a condition \lstinline|a<b| with two integer variables @a@ and @b@,
the boundary test objective is \lstinline|a-b+1==0|.
Since the exact boundary value can be unreachable, test engineers may want to get
sufficiently close to the boundary, up to a chosen distance value $N$.
This is the purpose of the criterion \textbf{LIMIT-N} of \LAnnot.
For example, for a condition \lstinline|a<b|, this requires to cover the
boundary test objective \lstinline|abs(a-b+1)<=N|.

\begin{figure}
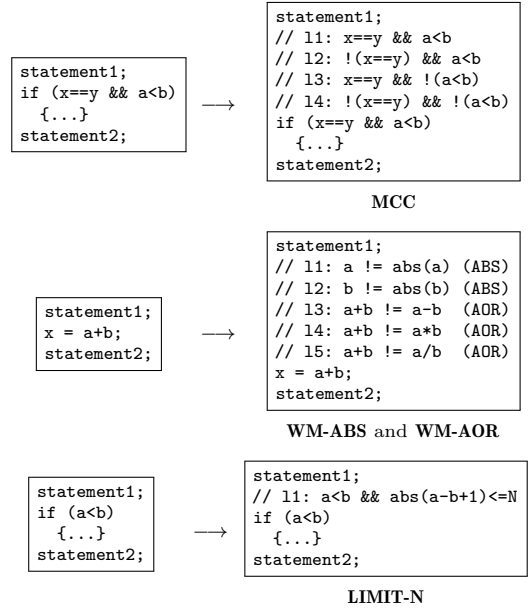

  \begin{footnotesize}
      \begin{minipage}{2.5cm}
        \begin{center}
\begin{boxedverbatim}
statement1;
if (x==y && a<b)
  {...}
statement2;
\end{boxedverbatim}
        \end{center}
      \end{minipage}
      \(\longrightarrow\)
      \begin{minipage}{4cm}
        \begin{center}
\begin{boxedverbatim}
statement1;
// l1: x==y && a<b
// l2: !(x==y) && a<b
// l3: x==y && !(a<b)
// l4: !(x==y) && !(a<b)
if (x==y && a<b)
  {...}
statement2;
\end{boxedverbatim}
~\\[0.5em]
\textbf{MCC}
        \end{center}
      \end{minipage}

\vspace{1em}

      \begin{minipage}{2.5cm}
        \begin{center}
\begin{boxedverbatim}
statement1;
x = a+b;
statement2;
\end{boxedverbatim}
        \end{center}
      \end{minipage}
      \(\longrightarrow\)
      \begin{minipage}{4cm}
        \begin{center}
\begin{boxedverbatim}
statement1;
// l1: a != abs(a) (ABS)
// l2: b != abs(b) (ABS)
// l3: a+b != a-b  (AOR)
// l4: a+b != a*b  (AOR)
// l5: a+b != a/b  (AOR)
x = a+b;
statement2;
\end{boxedverbatim}
~\\[0.5em]
\textbf{WM-ABS} and \textbf{WM-AOR}
        \end{center}
      \end{minipage}

\vspace{1em}

      \begin{minipage}{2.5cm}
        \begin{center}
\vspace{-2mm}
\begin{boxedverbatim}
statement1;
if (a<b)
  {...}
statement2;
\end{boxedverbatim}~\\[0.5em]
        \end{center}
      \end{minipage}
      \(\longrightarrow\)
      \begin{minipage}{4cm}
        \begin{center}
\begin{boxedverbatim}
statement1;
// l1: a<b && abs(a-b+1)<=N
if (a<b)
  {...}
statement2;
\end{boxedverbatim}
~\\[0.5em]
\textbf{LIMIT-N}
        \end{center}
      \end{minipage}
  \end{footnotesize}
\vspace{-3mm}
  \caption{Label annotation for some coverage criteria.}
\vspace{-7mm}
  \label{fig:labels}
\end{figure}

\figurename~\ref{fig:labels} shows label annotation for \textbf{MCC},
for weak mutations \textbf{WM-ABS} and \textbf{WM-AOR} with mutation operators ABS and AOR, and for  \textbf{LIMIT-N}.
Finally, custom labels can be added by test engineers when specific
test objectives are needed.

\vspace{-2mm}
\paragraph{Test generation for labels.}
Once coverage labels have been inserted into the program under test, they are transformed
in order to be used by the underlying DSE test generation tool.
Such a transformation (typically realized by program instrumentation)
determines a modified path exploration strategy and therefore represents
a key element of the resulting test generation technique.

\citet{BKC14} propose theoretical solutions to optimize test generation for labels.
First, they present two instrumentations,
called \emph{direct} and \emph{tight}.
Given a label $ℓ=(\loc,p)$ with location \loc and predicate $p$,
direct instrumentation replaces ℓ with a new branching instruction @if ($p$) {}@.
Tight instrumentation replaces ℓ with a non-deterministic choice leading to
a new assertion on $p$ and an exit: @if (nondet) { assert ($p$); exit (0); }@.

Direct instrumentation is shown by \citet{BKC14}
to exponentially increase the number of
paths in the instrumented program, contrary to the tight one.
Indeed, with direct instrumentation, for each path traversing a label location \loc in the initial program,
there are two paths going through \loc  (with $p$ being either true or false) in the instrumented program.
This multiplication of paths becomes even more significant
when some paths in the initial program traverse location
\loc several times (\eg because of loops or function calls).
As a consequence, a DSE tool generates a test for each path traversing a label location,
even if this label is already covered by a previously generated test.
To reduce such redundancy in test generation,
a first proposed optimization is tight instrumentation.
For each path leading to a label location, tight instrumentation
adds a unique path
that leads to the label condition verification and exits the program immediately after that,
otherwise the program ignores the label and continues.

The idea of the second optimization, called \emph{iterative label deletion} (ILD), is to replay tests as soon as they are generated to mark all labels that are covered during test execution.
This is used to prevent test generation from attempting to cover those labels anymore.
Indeed, the execution of a test generated to cover a given label ℓ may cover other labels,
for which it is no longer necessary to generate a test.
Tight instrumentation and iterative label deletion are expected to lead to a very efficient way for
dynamic symbolic execution to handle label coverage. This is confirmed by
\citet{BKC14} after a dedicated gray-box integration of these approaches
into a proprietary tool \PathCrawler~\cite{WMMR05}.
Until now, however, these optimizations have never been integrated into a publicly available testing tool,
nor have they been realized in a fully black-box manner.
In this work, we propose to leverage the genericity of this approach to
a renowned publicly available testing tool, and to do it in a fully black-box manner.
In doing so, we effectively extend the label-agnostic test generation tool \Klee to all coverage criteria that can be expressed with labels.
\vspace{-2mm}
\section{Motivating Examples}
\label{sec:motiv-ex}

\begin{lstlisting}[caption = {\texttt{power.c}},float=t,label={lst:power}]
int power (int X, int N) {
  int S = 1, Y = X, P = N;
  while (P >= 1) {
    if (P %
      P = P - 1;
      S = S * Y;
    }
    Y = Y * Y;
    P = P / 2;
  }
  return S;
}
\end{lstlisting}
%
%
The program @power@ of Listing~\ref{lst:power} takes two integer inputs @X@ and @N@ and computes $\mathtt{X}^\mathtt{N}$ when @N@ is non-negative%
\footnote{For simplicity, we ignore arithmetic overflows for this example.}.
In this program, each path corresponds to exactly one value of @N@. For example, the path entering the loop twice, the first time with an even value for @P@ and the second time with an odd value for @P@, only executes if @N=2@. When @N@ is bounded between 0 and a maximum value \(B\), with \(B>1\), \Klee explores \(B+1\) paths to always generate three tests\footnote{We use the option \texttt{--only-output-states-covering-new} so that \Klee only generates test cases covering yet uncovered code.}:
one with @N=0@, one with an even value for @N@ and one with an odd value for @N@. As one increases the value of \(B\), the exploration time grows, even if only the three same tests are produced in the end.

For this kind of programs consisting of a very simple control flow with a large set of different paths, \Klee spends a lot of time exploring all the paths: \num{8.3}\,s for $B=100$, 52\,s for $B=1000$, 553\,s for $B=5000$, to give a few examples.
However, only a few paths are necessary to cover basic criteria like instructions, decisions or conditions. We will show how the support of labels can drastically improve the exploration time of such programs while still producing a relevant test set of a small size.

As another example, consider the function @search@ of Listing~\ref{lst:search}. For a fixed maximal length \(L>1\) of @tab@, and @n@ assumed to be between 0 and \(L\), \Klee generates 3 tests covering all branches, for instance: (1) @n=0@, (2) @n=1@, @tab[0]=0@ and @val=0@, (3) @n=2@,  @tab[0]=42@,
@tab[1]=0@, and @val=0@. We observe that decisions and conditions are  covered, but the multiple condition @!res && i<n@ is never evaluated with the combination @!res@ and @!(i<n)@. Indeed, in the three tests generated, the searched value is found in the array. The execution exits the loop because @res=1@ and not because the end of the array is reached. Therefore, the test suite generated by \Klee does not satisfy multiple condition coverage.

\begin{lstlisting}[caption = {\texttt{search.c}},float=t,label={lst:search}]
int search (int *tab, int n, int val) {
  int res = 0, i = 0;
  while (!res && i < n) {
    if (tab[i] == val)
      res = 1;
    i++;
  }
  return res;
}
\end{lstlisting}

We observe that for
coverage criteria that are not subsumed by all-path (like multiple conditions, weak mutations, or condition limits), the test suites generated by \Klee may be incomplete. We will show how the support of labels helps to achieve a high coverage of such criteria with a reasonable overhead. %

\vspace{-2mm}
\section{Design of an Optimised Approach}
\label{sec:instru}

%

%

%
%
%
%
%
%
%
%
%
%

%
%
%
%
%
%

%
%
%
%
%
%
%

%
%

%

%
%
%
%
%
%
%
%
%
%
%
%
%
%

%
%

%
%
%
%

%

%
%
%
%
%
%
%

%
%
%
%

%

%

%

%
%
%
%
%

Before delineating our optimized approach for targeting labels in \Klee, we first describe a naive approach and illustrate the overall test generation process.
Along with informal notation for labels of the form @// lid: expr@
  used in the examples of \figurename~\ref{fig:labels},
we will use a macro \lstinline{cov_label(expr, id)} to denote labels in annotated C programs, where
\lstinline{id} is a unique integer identifier associated with the label,
and \lstinline{expr}  is the C expression of its predicate $p$.
The macro is practical to define the required behavior, which can vary between the different test generation approaches and for test replay.

\vspace{-2mm}
\subsection{%
  Naive Approach}
\label{sec:over-gener-proc}
\label{sec:naive-instr}

\begin{figure}[t]
  \footnotesize
  \centering
  \begin{minipage}{2.5cm}
    \begin{boxedverbatim}
statement1;
cov_label(expr, id)
statement2;
\end{boxedverbatim}
  \end{minipage}
  \hspace{0.5em}
  $\longrightarrow$
  \begin{minipage}{2cm}
    \includegraphics[width=2cm]{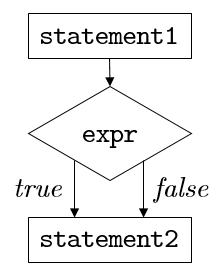}
  \end{minipage}
\vspace{-3mm}
  \caption{Naive (direct) instrumentation of labels}
\vspace{-5mm}
  \label{fig:naive}
\end{figure}

The naive approach to the problem corresponds to direct instrumentation (cf. Section~\ref{sec:coverage-labels}),
where each label
adds an additional branching instruction
whose both branches lead back to the next statement.
Instrumenting the program to achieve this
consists in replacing each label @cov_label(expr, id)@ by a new branching instruction @if (expr) {}@, as illustrated in \figurename~\ref{fig:naive}.
In this way, we effectively instruct \Klee to try to generate at least one test case that covers the empty branch, \ie where @expr@ holds and the label is covered.

In practice, \Klee currently does not handle empty code blocks (\eg inserted by a statement that cannot be optimized out by a C compiler, like @__asm__ volatile ("");@).
To circumvent this limitation, we implement the empty branch using a call to an external function @nop@ that returns immediately\footnote{This function is given in a separate library, and its implementation is therefore not subject to \Klee's scrutiny: it is only called when \Klee's concrete execution traverses this branch.}.

\paragraph{Measuring Coverage}

Executing \Klee on the instrumented program produces a set of test cases \(T\).
Measuring the coverage of \(T\) \wrt a given criterion boils down to replaying each test case from \(T\) and recording all covered labels.
We replay all generated test cases
using a specific instrumentation of the program under test.
A coverage status for each label (covered or uncovered) is recorded in a shared, persistent store that we denote by σ.

More concretely, we construct an executable version of the program called the \emph{replayer}, where each label @cov_label(expr, id)@ is replaced with a call to a function @set_covered@ guarded with @expr@, \ie
@if (expr) { set_covered (id); }@.
The role of @set_covered (id);@ is to register into σ that the label identified with @id@ has been covered.
After replaying every test case using the replayer,
the coverage achieved by \(T\) is recorded in σ and can be measured.

\paragraph{Runtime Errors}
\label{sec:rte}
Notice that, as mentioned in Section~\ref{sec:klee}, one of \Klee's goals is to produce test cases that
trigger runtime errors (RTEs).
Yet, we need to take special care regarding such test cases for the two following reasons.
First, our objective is \emph{not} to find test cases that trigger
RTEs: instead, we want to achieve label coverage.
If RTEs are detected, they should be reported, analyzed and fixed separately (\eg by adjusting the preconditions or fixing possible bugs in the code). This task is outside the scope of this work.
Second, RTEs actually correspond to \emph{undefined} behaviors,
hence, by nature, label coverage achieved after an undefined
behavior cannot be considered as firmly achieved.

We have thus chosen to keep only tests generated for complete paths, and to ignore tests with an .err file (for assertions or RTEs).
In practice, to correctly record label coverage,
we redefine the role of the @set_covered@ function used by the replayer,
and make it insert new label statuses into a temporary
buffer \(\tilde{σ}\).
Then, at the end of its execution, if it is successful (\ie the test execution terminates normally), the replayer commits the contents of \(\tilde{σ}\) into the store σ.
In this way, when a test case \(t\) triggers an RTE \emph{after} having covered a label ℓ, the replayer does not effectively update the status of ℓ in the persistent store σ, and the coverage induced by \(t\) is thus not accounted for: \(t\) is removed from \(T\) without compromising the consistency of the reported coverage measures.
This gives us the set of test cases \(T'\), with \(T' \subseteq T\).

\vspace{-2mm}
\subsection{Optimized Approach}

While it is capable of achieving label coverage in theory, several drawbacks cripple the naive approach delineated above in practice.
First, direct instrumentation induces an exponential explosion in the number of paths that must be explored.
Second, the constraints (\emph{path conditions})
that are accumulated during the symbolic execution of a path grow in
complexity each time the path traverses the branching instruction added for a label.
This can be observed in \figurename~\ref{fig:naive}, where every path reaching @statement2@ traverses the guard @if (expr)@.

\paragraph{Tight Instrumentation}
\label{sec:tight-instr}

In their quest for a more efficient test case generation for label coverage, \citet{BKC14} first designed tight instrumentation by observing that:
\begin{enumerate*}[(i)]
\item the expression @expr@ in @cov_label(expr, id)@ is only relevant to covering the label; and
\item the expression @!expr@ is irrelevant in any path.
\end{enumerate*}
Tight instrumentation ``cuts'' the branch that covers a label, and
prevents the propagation of @!expr@ into path conditions for longer
paths.
To design a practically applicable version of the general
idea of \citet{BKC14}, we guard the condition on @expr@
for a label @cov_label(expr, id)@
with a non-deterministic choice encoded using an additional symbolic variable whose \emph{unique} name is built using @id@.
Since it is never assigned anywhere in the program, this
variable is effectively an additional input used to simulate the
non-deterministic choice for the label: either the label is to be
considered, or it is ignored.

\Klee provides several primitives that allow us to achieve this.
First, we can force the generation of a test case with expression @e@ being {null} by using a statement @klee_assert (e);@.
Second, a statement @klee_silent_exit (0);@ stops any further exploration from its location, and does not generate any test case.
Finally, @klee_int ("varname")@ represents the value of a symbolic integer variable (of C type @int@) with name @varname@.
As a result, tight instrumentation can be achieved by replacing any label @cov_label(expr, id)@ with
\begin{lstlisting}
  if (NONDET (id)) {
    klee_assert (! (expr));
    klee_silent_exit (0);
  }
\end{lstlisting}
{\sloppy
where @NONDET(id)@ defines a symbolic integer variable as @klee_int ("nondet_" TOSTRING (id))@.
If the non-deterministic choice indicates that the label is to be considered (\ie @NONDET(id)@ holds), then
\Klee tries to generate a test with a failure of
@klee_assert ( !(expr))@ (that is, the label being covered). Otherwise the assert (and, therefore, the label) is ignored.
\par}

Contrary to the naive approach where the
set of test cases \(T\) consists of test cases generated by \Klee
for complete paths, in the optimized approach we are only interested
in keeping tests that violate a newly inserted assertion statement.
We achieve this straightforwardly by only considering tests for which a file with an \texttt{.assert.err} extension exists.
As a simple additional optimization, we insert a call @klee_silent_exit (0);@ at the very end of the tested program: in this way, we instruct \Klee to ignore every complete path that covers yet uncovered \emph{code}, but covers no label.

\paragraph{Iterative Label Deletion}

\emph{Iterative label deletion} (ILD) consists in preventing  symbolic execution from trying to cover a label that has already been covered.
Our optimized approach implements ILD by making use of an \emph{external} function @covered@\footnote{Similarly to
\lstinline|nop|, the function \lstinline|covered| is defined in a separate library.}
that, given the identifier @id@ of a label, returns a non-null integer if the label has already been covered.
This function simply queries the persistent store σ for the status of label @id@.

\begin{figure}%
  \includegraphics[width=\linewidth,trim=2mm 3mm 3mm 0mm,clip]{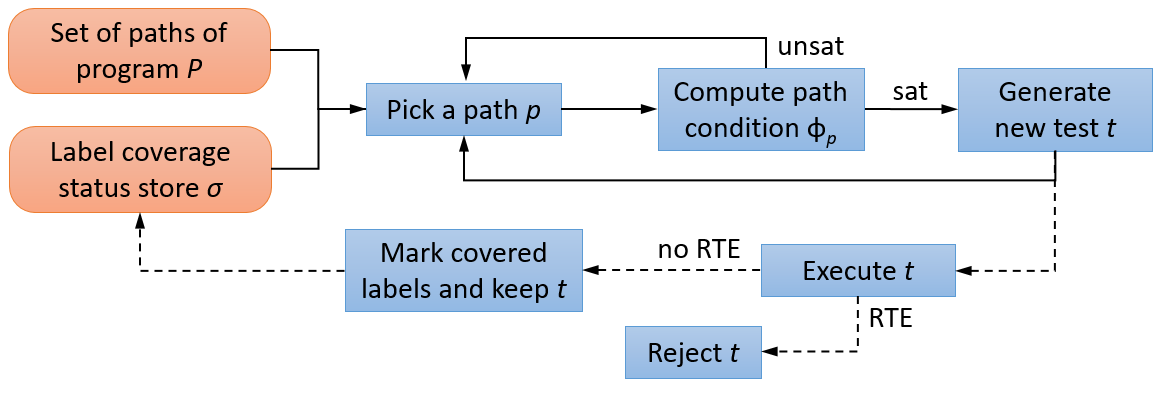}
\vspace{-3mm}
  \caption{Iterative label deletion}
\vspace{-4mm}
  \label{fig:ild}
\end{figure}

\begin{figure}
  \footnotesize
    \centering
    \begin{minipage}{2.5cm}
\begin{boxedverbatim}
statement1;
// label id: expr
statement2;
\end{boxedverbatim}
  \end{minipage}
$\longrightarrow$
\begin{minipage}{5cm}
\includegraphics[width=5cm]{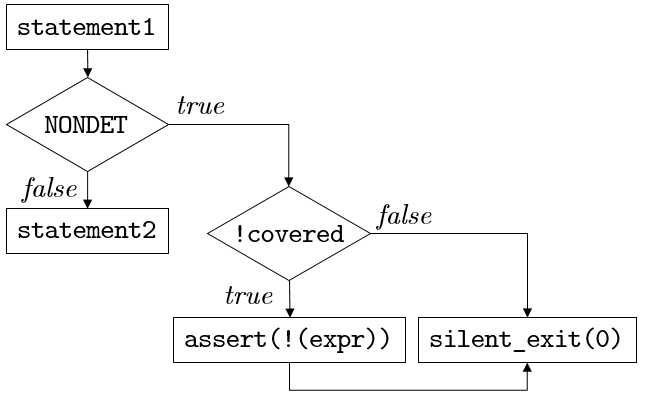}
\end{minipage}
\vspace{-3mm}
  \caption{Optimized instrumentation of labels}
\vspace{-5mm}
  \label{fig:optim}
\end{figure}

Our implementation of ILD is illustrated by \figurename~\ref{fig:ild},
where solid arrows indicate the main test generation flow and
dashed arrows show the replay steps performed in parallel.
Note that our implementation of ILD
is slightly different from
the theoretical approach of~\citet{BKC14}.
In fact, they propose to replay each test as soon as it is generated, while
the path exploration is suspended and waits for the results of the replay before continuing.
We do not record that a label is covered in σ as soon as the
corresponding assertion fails on a test case,
but only during its replay. Moreover, we only replay a test case when we detect that \Klee outputs a new test file that leads to a failure of the  assertion — \ie the name of its associated file is suffixed with \texttt{.assert.err}.
The path exploration of \Klee continues in parallel in the
meantime, as it is illustrated by the dashed lines in \figurename~\ref{fig:ild}.
This means that the result of subsequent calls to @covered@ may  depend on the timing of system-level operations.
An integration \textit{à la}~\citeauthor{BKC14} (where the path
exploration is suspended during the replay) may require altering the
source code of \Klee, to insert specific code to launch and wait for
a replayer sub-process as soon as a test case is generated.
This solution would delay the path exploration of \Klee by extra
waiting time and would not be compatible with our goal
to perform a black-box integration into \Klee.
  Our combination of tight instrumentation and ILD (illustrated in \figurename~\ref{fig:optim}) is achieved by replacing each @cov_label(expr, id)@ with:
\begin{lstlisting}
  if (NONDET (id)) {
    if (!covered (id))
      klee_assert (! (expr));
    klee_silent_exit (0);
  }
\end{lstlisting}
\noindent
By guarding the assertion with a call to @covered@, we cut any branch leading to an already covered label, and instruct \Klee to ignore the potentially complex expression of the label predicate.

An attentive reader will notice that this version is slightly different from the proposal of~\citeauthor{BKC14}~\cite[Fig.\,7]{BKC14}
that was sub\-optimal.
Indeed, in their proposal (and contrary to their declared intention),
in the case where a non-deterministic choice
indicates that the label must be considered but it is marked as already covered, the program does not exit
and symbolic execution continues the exploration of the rest of the program.
Hence the following branches can be explored in a redundant way,
when a non-deterministic choice is true and false.
This apparently small issue was rather tricky to find and to fix.
This confirms that it is important to validate a theoretical
approach by a working tool implementation.

\paragraph{Measuring Coverage}

An essential feature of the optimized approach
is that the code instrumented for the path exploration
in \Klee is not the same as for measuring the coverage of test cases.
Thanks to the intrumentation presented above,
during the path exploration, the condition of a label is considered only when necessary: at most once on a program path
and only if the label is not yet covered.
The coverage computation relies again on the replayer
described in Section~\ref{sec:over-gener-proc}.
Therefore, during the replay of a test case,
all labels are checked for being covered.

\vspace{-2mm}
\section{Evaluation}
\label{sec:experiments}

Thanks to adopting a black-box approach, the effort to implement our approach as a front-end for \Klee is limited.
Indeed, \red{our tool prototype \Kleelab} mostly consists of about \num{700} lines
of OCaml code, along with about \num{300} lines of C for
instrumentation macros and the library of external functions.
This prototype is publicly available\footnote{\url{https://github.com/OCamlPro/klee4labels} hosts the full source code of \Kleelab, along with several \Klee drivers that correspond to the examples of this section.
}.

To help us in evaluating our approach, we have also designed \red{an extended version of \Kleelab}, with a more advanced implementation of the label coverage store (whose optimized implementation is proprietary)\red{, an automated generation of test harnesses,} and further treatment of generated tests to produce readable C code for instance. \red{Nevertheless, the main optimization feature is the instrumentation of labels, which is the same in both versions of the tool. The rest of this section shows evaluation results based on the extended version of \Kleelab.}
We consider the following research questions:
\begin{enumerate}[RQ1,nosep,ref={\upshape RQ\arabic*}]
\item
  \label{rq:label-coverage}
  \textbf{Label coverage}: does the support for labels lead to a high coverage of labels by \Klee?
\item
  \label{rq:nb-tests}
  \textbf{Size of the test suite}: does the label-guided generation lead to test suites of reasonable sizes?
\item
  \label{rq:efficiency}
  \textbf{Efficiency}: what is the time overhead that is due to the exploration of \pclabels by \Klee?
\end{enumerate}

To answer these questions, we chose to
evaluate our optimized front-end for \Klee
on the same programs as in~\cite{BKC14}, plus a few demonstrative toy examples:
@tritype@ is a well-known program widely used for test generation;
@power@ is the program presented in Section~\ref{sec:motiv-ex};
@selection_sort@ is a classic implementation of a sorting algorithm;
@modulus@ comes from TestComp'22~\cite{TESTCOMP22};
@fourballs@ comes from \cite{PapadakisAST10};
@tcas@ and @replace@ come from the Siemens test suite~\cite{DER05};
@get_tag@ and @full_bad@ come from the Verisec benchmark~\cite{KHCL07};
and @gd@ comes from MediaBench~\cite{LPM97}.
Experiments were performed on a laptop computer with
{16}\,{GB}
of RAM and a
{1.8}\,{GHz}
Intel Core i7 CPU running GNU/Linux.
We used the latest development version of \Klee available\footnote{We built \Klee from its latest sources available at \url{https://github.com/klee/klee} on the the 25\textsuperscript{th}
  June 2022 (commit \texttt{667ce0f1}), with STP 2.3.3 as default solver.} with its default options%
, on our benchmark programs compiled to LLVM bitcode using Debian \clang version 14 (with default optimization settings).
The results\footnote{More detailed results can be found the long version of the paper on \url{https://doi.org/10.48550/arXiv.2211.14592}.}
are presented in \tablename~\ref{tab:stats}.

\begin{table*}
  \includegraphics[page=2,trim=0 5.7cm 0 0,clip]{table}%
  \caption{Selected experimental results, where `\timeout' denotes a timeout (set to 60\,s), and `\nan' denotes the absence of any usable measure obtained from \Klee after a timeout.%
  }
  \label{tab:stats}
\end{table*}
To evaluate our tool in terms of label coverage, test suite size and execution time, we ran it on the same programs with four different modes: a mode without label instrumentation, and three modes supporting labels in different ways.
In the first mode, called the \emph{ignore} mode in \tablename~\ref{tab:stats}, each label is replaced with an empty instruction block.
Thus, this mode corresponds to what \Klee does on its own, and is used as a witness.
The second mode implements the \emph{naive} instrumentation described in Section~\ref{sec:naive-instr}.
The third mode uses \emph{tight} instrumentation but does not replay test cases
as soon as they are generated, thus performing a full exploration of the instrumented program.
The last mode, called \emph{optim}, corresponds to our optimized implementation, with tight label instrumentation and iterative label deletion, and where the generated test cases are replayed in parallel.
In the three first modes, generated tests are replayed after the termination of \Klee in order to measure the reached coverage and to eliminate redundant tests in terms of coverage (with no guarantee of minimality).

We ran each mode on each program, with various coverage criteria: \textbf{DC}, \textbf{CC}, \textbf{MCC} and \textbf{WM}.
For two programs having sufficiently relevant conditions (@check@ and @gd@), we also used the \textbf{LIMIT} coverage criterion, which is \textbf{LIMIT-N} with $N=0$.
For each experiment, we limited the execution time of \Klee to %
60\,s.
Along with the total number of labels (\#labels) added to express  the selected criterion and the number of covered labels (\#covered), we recorded the following measures.
As reported by \Klee, we have the number of LLVM instructions (\#instr.) and the number of paths symbolically executed by \Klee (\#paths)\footnote{The number of paths is the sum of ``complete paths'' and ``partially completed paths'' reported by \Klee.}.
We report the number of tests kept after replay (\#tests, \ie \(|T'|\)) and the total number of tests generated by \Klee (gen, \ie \(|T|\)). They may differ a bit even in the optimized mode\footnote{Tests reaching RTEs are filtered out as explained in Section~\ref{sec:rte}. Moreover tests that only differ in the values assigned to non-deterministic inputs — the symbolic variables prefixed with \lstinline{"nondet_"} — are unified.}.
Finally, we only report the total execution time of \Klee as its execution time dominates every other computation of the whole front-end, including automated coverage label annotation by \LAnnot, and replays.
We denote timeouts with `\timeout', and also denote with `\nan' the absence of any usable measure obtained from \Klee after a timeout.

\paragraph{\ref{rq:label-coverage}.}
For simple criteria like \textbf{DC} or \textbf{CC}, the all-path coverage of Klee is sufficient to cover all labels.
However, when aiming at complex criteria like \textbf{MCC}, \textbf{WM} or \textbf{LIMIT}, all-path coverage misses some behaviors most of the time.
The need to help the exploration of the program with test objectives is prominent for these criteria.
Comparing the results of the instrumented modes to the ignore mode, the label coverage is improved in 16 out of 27 experiments.
With the optimized mode, the improvement spreads from 1,5\% up to 600\% (for @power@ with \textbf{WM}), with a median of 14\% and an average of 70\%.
This shows that label-guided test generation effectively leads to a better coverage of labels.

The coverage obtained in the ignore mode is on average 69\% of the chosen criterion, while reaching 80\% in the naive mode and 81\% in the tight and optimized modes.
In 10 out of 27 experiments, a coverage of more than 90\% is achieved with the optimized mode.
More accurate measures could be performed if we ruled out uncoverable labels (also called infeasible), since a coverage of 100\% is not always possible.
However, we have no immediate way to perform this analysis so far, and we leave the treatment of uncoverable labels as future work.

\paragraph{\ref{rq:nb-tests}.}
Since each instrumentation mode adds paths to the original program, and \Klee aims at covering all paths, the size of the generated test suite grows with the size of the path space.
For example, on @tcas@ with \textbf{WM}, as the number of explored paths grows from 44 (ignore), to \num{1484} (× 34, naive), to \num{2598} (× 59, tight), the number of generated tests grows from 18 (ignore), to 37 (× 2, naive), to 52 (× \num{2.9}, tight).
On some examples, the size of the test suites generated by the naive or the tight modes grows up to 7 (@checkutf8@ or @tritype@ with \textbf{WM}) and even 10 times (@fourballs@ with \textbf{WM}).

On the contrary, with the optimized mode, the size of the generated test suite (\ie \(|T|\)) is better controlled.
For the examples mentioned above, the optimized mode generates 18 tests for @tcas@ (× 1), 56 tests for @checkutf8@ (× \num{2.4}), 22 tests for @tritype@ (× \num{1.6}), 7 tests for @fourballs@ (× \num{1.8}).
This grows up to × 4 for @power@ with \textbf{WM}.
Interestingly, the optimized mode does not lead to a larger test suite than the ignore mode in 17 cases out of 27, and it leads to a strictly smaller one in 8
cases, for a resulting coverage that is at least equal.
The iterative deletion of labels that reduces the exploration to only yet uncovered labels proves to drastically limit the growth of the generated test suite.

As explained above, the generated test suite is filtered to eliminate tests reaching RTEs, as well as redundant tests that \Klee may generate due to non-deterministic inputs. Therefore the number of tests actually kept in the resulting test suite (\ie \(|T'|\)) is even smaller. The final test suite in the optimized mode is on average \num{0.98} times smaller than the one obtained with the ignore mode, and strictly smaller in 14 cases out of 27.

\paragraph{\ref{rq:efficiency}.}
The exploration time of \Klee grows with the size of the path space, and as explained above, the path space grows with each instrumentation.
This leads to an overhead of × \num{6.3} on average\footnote{We only compare experiments ending before the timeout.} for the naive mode (up to × 40 on @tcas@ with \textbf{WM}), and of × \num{8.8} on average for the tight mode (up to × 50 on @fourballs@ with \textbf{WM}).
On the contrary, for the optimized mode, the overhead is better controlled, with an average overhead of × \num{4.6} and a maximum of × 30 for @replace@ in \textbf{WM}.

On some examples, the optimized mode is even faster than the ignore mode: that is the case on programs with a lot of different paths but simple control-flow, like @power@, @selection_sort@ and @modulus@. For example on @modulus@, the optimized mode ends after %
\num{1.2}\,s,
while the ignore mode times out. On this kind of programs, for criteria like \textbf{DC}, \textbf{CC} or \textbf{MCC}, targeting labels drastically decreases the exploration time of \Klee, for the same achieved coverage.

Interestingly, the overhead for the optimized mode drops down to a maximum × \num{2.2} when the chosen criterion can be fully satisfied.
We noticed that \Klee spends a lot of time on uncoverable labels.
In fact, given a coverable label, either it is covered by the test specifically produced for it, or it is covered by executing another test, produced for another label.
In the case of an uncoverable label, the latter cannot happen, so \Klee needs to explore all the paths leading to this label, but finally fails to produce any test for it.
This loss of time is particularly noticeable in the examples where all coverable labels are covered in less than
60\,s
(sometimes in the first ten seconds), but the timeout is reached trying to cover the uncoverable ones (\eg @get_tag@ with \textbf{MCC} and \textbf{WM}).

\vspace{-2mm}
\section{Related Work}
\label{sec:related}

\paragraph{Test generation for labels.}

Support for labels was initially implemented inside the dynamic
symbolic execution tool
\PathCrawler~\cite{BKC14,BNMD21}. This tool supports
several control-flow coverage criteria from instruction to all-path
coverage, and a bounded version of all-path named $k$-path coverage~\cite{WMMR05}.
\PathCrawler's depth-first search algorithm was extended to support
labels in such a way that iterative label deletion is intimately
interleaved with symbolic execution. 
The experiments show the
efficiency of the approach with respect to the standard algorithm,
allowing for a better label coverage (from 3\% up to 39\%) for only a
slight overhead (median overhead of 37\%, average overhead of
115\%)~\cite{BNMD21}.

Our support of labels for \Klee follows a black-box approach,
therefore the efficiency of the two tools cannot be directly
compared. In particular, our instrumentation is totally independent of
\Klee's search algorithm, while the search algorithm of \PathCrawler
was significantly modified to support label coverage.
Moreover, in our tool, the replay of produced test cases is done using
the replay function provided by the API of \Klee, which launches the
executable from its very beginning. This induces an additional cost at
each test execution, while this cost is minimized in the integration to
\PathCrawler.
As the latter tool is not publicly
available, we were not
able to compare its efficiency to that of our implementation.
Based on the available evaluation results~\cite{BKC14},
we achieved comparable results in terms of
achieved coverage as well as execution time.

The \PathCrawler extension to labels has been used in industrial
experimentations~\cite{BKMMW18,LMC19}. The authors designed a unit
test generation tool that combines several formal and non-formal test
generation algorithms to target the MC/DC criterion. The C program is
first annotated with coverage labels for this criterion, then several
tools based on different techniques are called. After eliminating
infeasible labels, they generate a test suite targetting the remaining
labels. The tool uses a genetic algorithm, CBMC~\cite{CDE08}, and
\PathCrawler sequentially, to finally proceed to a test suite
optimization phase. The authors achieve a coverage of 99\% of MC/DC in
about half an hour, on a case study of 82,000 lines of C code with
integer data.
\red{This work gives interesting directions for improving
  efficiency and scalability.}
In particular, the detection of
infeasible labels prior to test generation, and the collaboration
between different tools, are the key to the scalability of the
approach.

\paragraph{DSE enhanced with coverage criteria.}

Several work extended white-box testing techniques and tools to coverage
criteria, for different programming languages. For example, in the
\textsc{Apex}~\cite{PXTH10,JFTH13} tool, which is an extension of
Pex~\citep{TH08}, coverage criteria like mutation or condition
boundaries are added as constraints in the path conditions computed by
the dynamic symbolic execution. Another extension of Pex for mutation
testing, PexMutator~\cite{ZXZTHM10}, instruments the program under
test with weak-mutant-killing constraints. Each of these constraints
is wrapped as a conditional statement, and Pex is then called on the
resulting program. This approach compares to what we call  naive
instrumentation in Section~\ref{sec:naive-instr}. Another comparable approach is
adopted for white-box testing for Java by~\citet{PM11}, where weak
mutations are encoded as new branching conditions in the source code,
of the form \textit{original expression $\neq$ mutated expression}.
None of these works tackle the path explosion problem caused by the
addition of branching conditions.

\paragraph{Assertion and RTE checking.}

Besides traditional coverage criteria, users often want to test given
properties of their programs, like the absence of certain runtime
errors and the satisfaction of assertions. This is the aim of
assertion-based testing~\cite{CKGJ12,GLM08,KA96}, where the analysis
of the program is reduced to the scope of the targeted property.
\citet{GLM08}, in particular, uses dynamic symbolic execution to compute
the set of all the paths reaching the targeted property: if none
of the resulting path conditions is satisfiable, the
unreachability of the property is proven; otherwise, a counterexample gives a
test case.

In a similar way, CBMC~\cite{CKL04} performs bounded model-checking on
a C source code, in order to check for reachability properties
(hand-written or known runtime errors). The tool is also able to
perform test case generation along different coverage criteria,
automatically encoded as reachability properties in the
code~\cite{AGNPS10}. This instrumentation of the code with extra
assertions to encode criteria makes it very close to our annotation
with labels. However, since model-checking does not explore
program paths one by one, no extra instrumentation is needed and the
BMC algorithm is very efficient when treating these additional
assertions.

\vspace{-2mm}
\section{Conclusion and Future Work}
\label{sec:conclusion}

A larger support of various coverage criteria in
test generation tools remains a challenging research topic.
Coverage labels offer a unified framework for specifying coverage criteria in a generic way.
However, labels are still not supported in popular
test generation tools.
Inspired by a previous theoretical proposal
for an efficient test generation technique
for labels by \citet{BKC14},
in this work we show how to integrate a dedicated support for labels
into \Klee~\cite{CDE08}, a popular open-source test
generation tool based on dynamic symbolic execution.
We perform a lightweight black-box integration,
which does not need to modify the underlying test generation strategy
and can therefore directly benefit of various strategies and
optimizations of the tool.
We expect that this work will facilitate and guide
a lightweight integration of this technique
into other similar tools.
Experiments with our version of the tool on several benchmarks confirm the benefits of the proposed approach.
In particular, our approach efficiently achieves basic criteria
while generating fewer and more targeted test cases
than when \Klee is used directly.
On more advanced criteria like multiple conditions, weak mutations,
and condition limits, our approach is capable of efficiently
achieving high coverage with a reasonable overhead.

Future work includes a large industrial evaluation
of the proposed approach on real-life code.
Detecting infeasible objectives before running test generation
is another promising perspective,
since it can avoid the waste of time and effort of
test generation tools
trying to cover test objectives that cannot be covered.
It can rely on tools like CBMC~\cite{CKL04}
and the LUncov module of LTest~\citep{BNMD21}.
Another future work direction is
a further extension of the proposed approach to
other criteria not expressible with labels~\cite{MDBKP17,MKPL20}.
Finally, integration of the proposed approach into other
tools, based on dynamic symbolic execution \red{or other test
generation techniques like fuzzing}, can be an interesting
extension for these tools.

\vspace{-2mm}

\bibliographystyle{ACM-Reference-Format}
\bibliography{biblio}

\clearpage
\clearpage
\section*{Appendix}
\label{sec:appendix}
We provide on the following pages a complete table with the
evaluation results for all our executions on the benchmark programs.
\clearpage
\onecolumn{%
  \centering
  \smaller
  \renewcommand{\arraystretch}{.92}%
\begin{longtable}{@{}lllrrrr@{}}
\toprule
 & \criterionName (\#labels) & measure & ignore & naive & tight & optim \\
\midrule
\endfirsthead
\toprule
 & \criterionName (\#labels) & measure & ignore & naive & tight & optim \\
\midrule
\endhead
\midrule
\multicolumn{7}{r}{Continued on next page} \\
\midrule
\endfoot
\bottomrule
\endlastfoot
\multirow[t]{24}{*}{\textsf{power} {(18 loc)}} & \multirow[t]{6}{*}{DC (4)} & \#covered & \num[zero-decimal-to-integer,scientific-notation=false]{4} & \num[zero-decimal-to-integer,scientific-notation=false]{4} & \num[zero-decimal-to-integer,scientific-notation=false]{4} & \num[zero-decimal-to-integer,scientific-notation=false]{4} \\
\textsf{} &  & \#instr. & \num[zero-decimal-to-integer,scientific-notation=false]{4692.0} & \num[zero-decimal-to-integer,scientific-notation=false]{8230.0} & \num[zero-decimal-to-integer,scientific-notation=false]{16279.0} & \num[zero-decimal-to-integer,scientific-notation=false]{2192.0} \\
\textsf{} &  & \#completed paths & \num[zero-decimal-to-integer,scientific-notation=false]{101.0} & \num[zero-decimal-to-integer,scientific-notation=false]{101.0} & \num[zero-decimal-to-integer,scientific-notation=false]{0.0} & \num[zero-decimal-to-integer,scientific-notation=false]{0.0} \\
\textsf{} &  & \#partially completed paths & \num[zero-decimal-to-integer,scientific-notation=false]{0.0} & \num[zero-decimal-to-integer,scientific-notation=false]{0.0} & \num[zero-decimal-to-integer,scientific-notation=false]{829.0} & \num[zero-decimal-to-integer,scientific-notation=false]{111.0} \\
\textsf{} &  & \#tests /gen & 3 /3 & 3 /3 & 3 /4 & 2 /4 \\
\textsf{} &  & time (s) & 8.3 & 7.6 & 7.9 & 1.2 \\
\cline{2-7}
\textsf{} & \multirow[t]{6}{*}{CC (4)} & \#covered & \num[zero-decimal-to-integer,scientific-notation=false]{4} & \num[zero-decimal-to-integer,scientific-notation=false]{4} & \num[zero-decimal-to-integer,scientific-notation=false]{4} & \num[zero-decimal-to-integer,scientific-notation=false]{4} \\
\textsf{} &  & \#instr. & \num[zero-decimal-to-integer,scientific-notation=false]{4692.0} & \num[zero-decimal-to-integer,scientific-notation=false]{8230.0} & \num[zero-decimal-to-integer,scientific-notation=false]{16279.0} & \num[zero-decimal-to-integer,scientific-notation=false]{2181.0} \\
\end{longtable}
}\twocolumn
 
\end{document}